\documentstyle[floats,twocolumn,aps,psfig,prl]{revtex}
\begin{document}
\draft
\twocolumn[\hsize\textwidth\columnwidth\hsize\csname @twocolumnfalse\endcsname

\title{Conduction channels of superconducting quantum point contacts}

\author{E. Scheer, J.C. Cuevas, A. Levy Yeyati, A. Mart\'\i n-Rodero, P. Joyez, 
M.H. Devoret, D. Esteve and C. Urbina}
\address{$^1$Physikalisches Institut, Universit\"at Karlsruhe,
76128 Karlsruhe, Germany; $^2$Departamento de F\'\i sica Te\'orica de la Materia Condensada C-V, 
Universidad Aut\'onoma de Madrid, 28049 Madrid, Spain; $^3$Service de Physique de l'Etat Condens\'e, Commissariat \`a
l'Energie Atomique, Saclay, 91191 Gif-sur-Yvette Cedex, France}

\maketitle

\begin{abstract}

Atomic quantum point contacts accommodate a small number of
conduction channels. Their number $N$ and transmission coefficients  
$\left\{T_{\mathrm n}\right\}$ can be determined by analyzing the 
subgap structure due to multiple Andreev reflections in the current-voltage ($IV$) 
characteristics in the superconducting
state. 
With the help of mechanically controllable break-junctions we have
produced Al contacts consisting of a small number of atoms. In the smallest stable contacts, 
usually three channels contribute to the transport. 
We show here that the channel ensemble $\left\{ T_{\mathrm n}\right\}$ of few atom contacts remains 
unchanged up to temperatures and magnetic fields approaching the critical 
temperature and the critical field, respectively, giving experimental evidence for the prediction that the
conduction channels are the same in the normal and in the superconducting state. 
\end{abstract}

\vskip2pc]

\section{Introduction}
 An atomic size contact
between two metallic electrodes can accommodate only a small number of conduction channels. 
The contact is thus fully described by a set 
$\left\{ T_{\mathrm n}\right\} =\left\{T_1,T_2,...T_{\mathrm N}\right\} $ of transmission
coefficients which depends both on the
chemical properties of the atoms forming the contact and on their
geometrical arrangement. Experimentally,
contacts consisting of even a single atom have been obtained using both
scanning tunneling microscope and break-junction techniques \cite{jancuracao}. The total 
transmission $D$=\thinspace $\sum\limits_{n=1}^N T_{\mathrm n}$ of a contact is deduced 
from its
conductance $G$ measured in the normal state, using the Landauer formula $G~=~G_0D$ where
$G_0=2e^2/h$ is the conductance quantum \cite{landauer}.

Experiments on a large ensemble of metallic contacts have
demonstrated the {\it statistical} tendency of atomic-size contacts to adopt configurations leading
to some preferred values of conductance. The actual preferred values depend on the metal and on the
experimental conditions. However, for many metals, and in particular 'simple' ones (like Na, Au...)
which in bulk are good 'free electrons' metals, the smallest contacts have a conductance $G$ close
to $G_0$ \cite{jancuracao,olesen,krans95,costa-krämer}. Statistical examinations of Al point
contacts at low temperatures yield preferred values of conductance at $G = 0.8~G_0, 1.9~G_0,
3.2~G_0$ and $4.5~G_0$ \cite{yansoncomment}, indicating that single-atom contacts of Al have a
typical conductance
slightly below the conductance quantum. Does this mean that the single-atom contacts correspond 
to a single, highly transmitted channel $(T = 0.8)$? This question cannot
be answered solely by conductance measurements which provide no information
about the number or transmissions of the individual channels.

However, it has been shown that the {\it %
full} set $\left\{ T_{\mathrm n}\right\} $ is amenable to measurement in the case
of superconducting materials \cite{aluprl} by quantitative comparison of the measured 
current-voltage ($IV$
characteristics) with the theory of multiple Andreev reflection (MAR) for a single channel BCS
superconducting contact with arbitrary transmission $T$, developed by 
several groups for zero temperature $\Theta=0$ and zero magnetic field $H=0$
\cite{arnold87,averin95,cuevas,shumeiko}. 
Although the typical conductance of single-atom contacts of Al ($G \simeq 0.8~G_0$) is smaller 
than the maximum possible conductance for
one channel, three channels with transmissions such that  
$T_{\mathrm 1}+T_{\mathrm 2}+T_{\mathrm 3} \simeq 0.8$ have been found \cite{aluprl}.

Moreover, there exist other physical properties which are not linear with respect to
$\{T_{\mathrm n}\} $ as e.g. shot noise \cite{helko}, 
conductance fluctuations \cite{ludophcf}, and thermopower 
\cite{ludophtp}, which also give information about the
$\left\{ T_{\mathrm n}\right\} $ of a contact. Although it is not possible to determine  the full
set of transmission coefficients with these
properties, certain moments of the distribution and in particular the presence or absence of partially open
channels can be detected. Recent experiments have shown that normal 
atomic contacts of Al with conductance close to $G_0$
contain incompletely open channels \cite{helko} in agreement with the findings in the 
superconducting
state \cite{aluprl}.  
 
In previous work we have shown \cite{carlosprl,scheernature} how the conduction channels of 
metallic contacts can be constructed from the valence orbitals of the material under investigation. 
In the case of
single-atom contacts the channels are determined by the valence orbitals of the
central atom and its local environment. In particular for Al the channels arise
from the contributions of the $s$ and $p$ valence bands. 
To the best of our knowledge it has never been observed in contacts of 
multivalent metals that a single channel arrives at its saturation value of $T = 1$ before at least a second
one had opened. 
 Single-atom contacts of the monovalent metal Au transmit one single channel with a 
transmission $0 < T \leq 1$ depending on the particular realization of the contact 
 \cite{scheernature,aupaper}.

From the theoretical point of view no difference between the normal and superconducting
states is expected, because ($i$) according to the BCS
theory \cite{BCS} the electronic wave functions themselves are not altered when entering the
superconducting state, but only their occupation and ($ii$) MAR preserves electron-hole symmetry and therefore does not mix 
channels \cite{beenakkerprb,averinprbrc}. 

Experimental evidence for the equivalence of the normal and superconducting channels can be 
gained by tracing the evolution of the $IV$ curves from the superconducting to the normal state
in an external magnetic field and/or higher temperatures and comparing them to the recent
calculations by Cuevas {\sl et al.} of MAR in single channel contacts
at finite 
temperatures \cite{carlosdiss} and including pair breaking due to magnetic impurities or 
 magnetic field \cite{carlosunpub}. 
 
 We show here that the channel ensemble 
 $\left\{ T_{\mathrm n}\right\}$ of few atom contacts remains 
unchanged when suppressing the superconducting transport properties gradually by raising the
temperature or the magnetic field up to temperatures and magnetic fields approaching the critical 
temperature and the critical field, respectively. 
 Although it is
not possible to measure the full channel ensemble above the critical temperature or field,
respectively, no abrupt change is to be expected since the phase transition (as a function of
temperature) is of second order.
Because the determination of the
channel ensemble relies on the quantitative agreement between the theory and the 
experimental $IV$s, we concentrate here on the case of Al point contacts since we expect for this
material the BCS theory to fully apply.

\section{Transport through a superconducting quantum point contact}

\begin{figure}[t!]
\centerline{\psfig{file=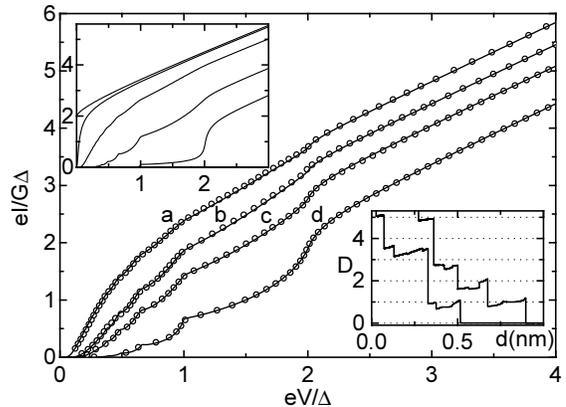,width=0.85\linewidth}}
\caption{Measured $IV$ curves (symbols) of four
different atomic contacts with $G\simeq 0.9~G_0$ at $\Theta \leq$ 50 mK and best numerical
fits (lines). The $\{ T_{\mathrm n}\} $ and total transmissions $D$ obtained from the fits are: 
(a) $\{0.900, 0.108\}$, $D$ = 1.008; 
(b) $\{0.802, 0.074\}$, $D$ = 0.876; 
(c) $\{0.747, 0.168, 0.036\}$, $D$ = 0.951;
(d) $\{0.519, 0.253, 0.106\}$, $D$ = 0.878. 
Current and voltage are in reduced units, the
current axis is normalized with respect to the total conductance measured by 
the slope of the
$IV$ at high voltages $eV > 5\Delta$. Not all measured data points 
are shown.
The measured gap was $\Delta /e~=~(184~\pm~2)~\mu $V. Left inset: 
Theoretical {\sl IV}s for a
single channel superconducting contact for different values of 
$T$ (from bottom to top:
0.1, 0.6, 0.9, 0.99, 1) after \protect\cite{cuevas}. 
Right inset: typical total transmission
traces measured as a function of the electrode distance 
$d$ at $eV > 5~\Delta $, while opening 
the contact at around 6 pm/s. The $d$ axis has been calibrated 
by the exponential behavior in the
tunnel regime.}
\label{aluivs}
\end{figure}

The upper left
inset of Fig.~\ref{aluivs} shows the theoretical $IV$s by Cuevas {\sl et al.} \cite{cuevas} for zero 
temperature $\Theta = 0$ and zero field $H = 0$. A precise
determination of the channel content of any superconducting contact is 
possible making use of the fact that the total current $I(V)$ results from the
contributions of $N$ independent channels: 
\begin{equation}
I(V)=\sum\limits_{n=1}^Ni(V,T_n). 
\label{sum}
\end{equation}
This equation is valid as long as the scattering matrix whose eigenvalues are given by the
transmission coefficients is unitary, i.e. the scattering is time independent. The $i(V,T)$ curves present a series of sharp current steps at voltage values 
$V$~=$~2\Delta /me$, where $m$ is a positive integer and $\Delta $ is the
superconducting gap. Each one of these steps corresponds to a different
microscopic process of charge transfer setting in. For example, the
well-known non-linearity at $eV=2\Delta $ arises when one electronic charge $%
(m=1)$ is transferred thus creating two quasiparticles. 
The energy $eV$ delivered by the voltage source must be
larger than the energy $2\Delta $ needed to create the two excitations. The
common phenomenon behind the other steps is multiple Andreev reflection
(MAR) of quasiparticles between the two superconducting reservoirs 
\cite{klapwijk,hurd}. The order $m=2,3,...,$ of a step corresponds to the number
of electronic charges transferred in the underlying MAR process.
Energy conservation imposes the threshold $meV \geq 2\Delta $ for each process.
For low transmission, the contribution to the current arising from the
process of order $m$ scales as $T^m$. The contributions of all processes sum up
to the so-called "excess current" the value of which can be determined by extrapolating
the linear part of the $IV$s well above the gap $v > 5 \Delta$ down to zero voltage. 
As the transmission of the channel
rises from 0 to 1, the higher order processes grow stronger and the 
current increases progressively. The ensemble of steps is called "subharmonic gap
structure", which was in fact discovered experimentally \cite{burstein},
has been extensively studied in superconducting weak links and tunnel
junctions with a very large number of channels \cite{flensberg89,kleinsasser}. 

\section{Experimental techniques} 

In order to infer $\left\{ T_{\mathrm n}\right\} $ from the $IV$s, very stable
atomic-size contacts are required. For this purpose we have used
micro-fabricated mechanically controllable break-junctions \cite{jansaclay}. 
Our samples are 2\,$\mu $m long, 200 nm thick suspended
 microbridges, with a $100$\,${\rm nm}\times 100$\,${\rm %
nm}$ constriction in the middle (cf.~Fig.~\ref{setup}). The bridge is broken at the
constriction by controlled bending of the elastic substrate mounted on a
three-point bending mechanism. A differential screw (100\,$\mu $m
pitch) driven by a dc-motor through a series of reduction gear boxes,
controls the motion of the pushing rod that bends the substrate (Fig.~\ref{setup}).

\begin{figure}[t!]
\caption{Three point bending mechanism. The distance between the two counter-supports is
12\,mm, and the substrate is 0.3\,mm thick. The
micrograph shows a suspended Al microbridge. The insulating
polyimide layer was etched to free the bridge from the substrate. The third panel displays the
wiring of the experimental setup.}
\label{setup}
\end{figure}

The geometry of the bending mechanism is such that a 1\,$\mu $m
displacement of the rod results in a relative motion of the two anchor
points of the bridge of around 0.2~nm. This was verified using the exponential dependence of the
conductance on the interelectrode distance in the tunnel regime. This very strong dependence was
used to calibrate the distance axis to an accuracy of about 20 \%. The bending mechanism is anchored 
to the mixing chamber of a
dilution refrigerator within a metallic box shielding microwave frequencies. 
The bridges are broken at low temperature and under cryogenic vacuum to avoid contamination.

The $IV$ characteristics are measured by voltage biasing with $U = U_{\mathrm dc}$ 
the sample in series with a calibrated
resistor $R_{\mathrm s} = 102.6\,{\mathrm k}\Omega$ and measuring the voltage drop across the sample 
(giving the $V$ signal) and the voltage drop $V_{\mathrm S} = IR_{\mathrm s}$ across 
$R_{\mathrm s}$ 
(giving the $I$ signal)
 via two low-noise differential preamplifiers. The differential conductance is measured by biasing
 with $U = U_{\mathrm dc}+U_{\mathrm 1}\cos(2\pi f t)$ using
a lock-in technique at low frequency $f < 200~{\mathrm Hz}$. 
All lines connecting the sample to the room 
temperature electronics are carefully filtered at microwaves frequencies by
a combination of lossy shielded cables \cite{glattli}, and microfabricated
cryogenic filters \cite{vion}.
 The cryostat is equipped with a superconducting solenoid allowing to control the field 
 $\mu_0 H$ at the position of the sample within  
0.05 mT. After having applied a magnetic field and before taking new $H = 0$ data we demagnetize 
carefully the solenoid. The temperature is monitored 
by a calibrated resistance thermometer thermally anchored to the shielding box. The 
absolute accuracy of the temperature measurement is about 5\%.

\section{Determination of the channel transmissions}

Pushing on the
substrate leads to a controlled opening of the contact, while the sample is
maintained at $\Theta < 100~{\mathrm mK}$. As found in previous experiments at higher temperatures,
the conductance $G$ decreases in steps of the order of $G_0$, their exact
sequence changing from opening to opening (see right inset of Fig.~\ref{aluivs}).
The last conductance value before the contact breaks is usually between 0.5
and 1.5~$G_0$. 

Figure~\ref{aluivs} shows four examples of $IV$s obtained  at 
$\Theta < 50~{\mathrm mK}$ on last plateaux of two different Al samples just before breaking the contact
and entering the tunnel regime. The curves differ markedly eventhough they
correspond to contacts having the same conductance of about $G \simeq 0.9~G_0$ within 10\%.
The existence of $IV$s with the same conductance but different subgap structure implies the
presence of more than one channel without further analysis. In particular, the examples shown here
demonstrate that although they would correspond to the first maximum in the conductance histogram
\cite{yansoncomment}, they do not transmit a
single channel, but at least two, with a variety of transmissions. 

In Fig.~\ref{aluivs} we also show the best least-square fits obtained using the
numerical results of the $\Theta~=~0$ theory of Cuevas {\sl et al.} \cite{cuevas}. The fitting
procedure decomposes the total current into the
contributions of eight independent channels. Channels found with transmissions
lower than 1\% of the total transmission were neglected.  When $N\leq $\,$3$, this
fitting procedure allows the determination of each $T_{\mathrm n}$ with an
accuracy of 1\% of the total transmission $D$. For contacts containing more channels only
the 2 or 3 dominant channels (depending on their absolute value) can be
extracted with that accuracy. Details of the fitting procedure are published in \cite{aluprl}.

\section{IVs of Al point contacts at higher temperatures}

\begin{figure}[t]
\centerline{\psfig{file=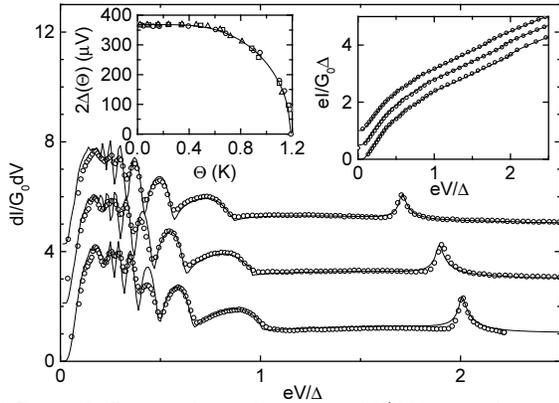,width=0.85\linewidth}}
\caption{Differential conductance ${\mathrm d}I/{\mathrm d}V$ as a function of voltage measured at 
$\Theta =$
47 mK (bottom),~640 mK (middle), and 810 mK (top) for a contact with total conductance $D =
1.008$. The two upper curves are offset by $2$ each. The solid lines are fits
using the same transmission ensemble $\left\{ T_1=0.900, T_2=0.108\right\}$ for all three
temperatures. Right inset: corresponding $IV$s offset by $0.5$ each. Left
inset: position of the $m=1$ maximum in the ${\mathrm d}I/{\mathrm d}V$ curves. Data has been taken
on three different contacts with $G$ varying from 1 to 2 $G_0$.
The line is the temperature dependence of the superconducting gap according to BCS theory for
$\Delta(\Theta=0)= 184 \mu {\mathrm V}$.}
\label{tdep}
\end{figure}

The right inset of Fig.~\ref{tdep} displays the evolution of the $IV$ of contact (a) 
from Fig.~\ref{aluivs} for three different temperatures below the critical temperature
$\Theta_{\mathrm c}\,=\,1.21\,{\mathrm K}$ (each trace is offset 
by $0.5$ for clarity). When the temperature is increased the subgap 
structure is slightly smeared out due 
to the thermal activation of quasiparticles, and the
position of the current steps is shifted to smaller voltages due to the reduction of the 
superconducting gap. 
 Although the $IV$s are very smooth due to the dominance of an
almost perfectly open channel, up to eight MAR processes are distinguishable in the
${\mathrm d}I/{\mathrm d}V$. The solid lines are calculated with the same set of 
transmissions $\left\{ T_1=0.900, T_2=0.108\right\}$ for the temperatures indicated in the caption
using the BCS dependence of the superconducting 
gap and the Fermi function for the respective
temperature \cite{carlosdiss,hurd}. The quality of the fit does not vary with temperature. 
\begin{figure}[t]
\centerline{\psfig{file=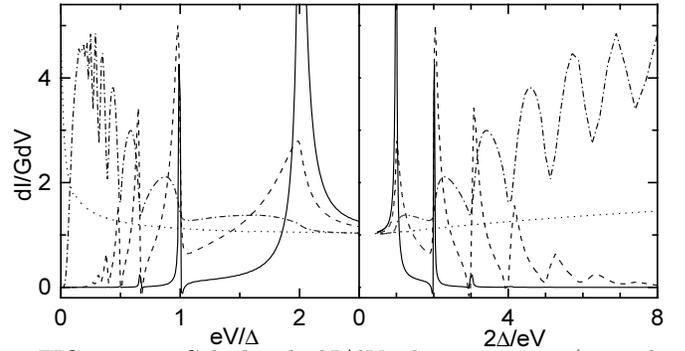,width=1.00\linewidth}}
\caption{ 
Calculated ${\mathrm d}I/{\mathrm d}V$ characteristics (normalized to the conductance) of a single 
channel quantum point contact for
transmission 0.1 (solid), 0.6 (dashed), 0.9 (dash-dotted), 1.0 (dotted) as a function of $eV/\Delta$ (left
panel) and $m^\prime =2\Delta /eV$ (right panel). The maximum of the curve for $T = 0.1$ arrives 
at $12.6 G$. The amplitude, the shape and the exact position of 
the individual maxima depend on the
transmission. Only for small $T$ the maxima are close to $V =2\Delta/me$, for higher transmission
they are shifted to lower voltages. The subgap structure is most pronounced for intermediate $T$. 
}\label{theorydidv}
\end{figure}

In order to further interpret the data, we
plot in Fig.~\ref{theorydidv} the theoretical zero temperature single channel 
${\mathrm d}I/{\mathrm d}V$s for the same transmissions as in the inset in Fig.~\ref{aluivs} 
in two different manners. In the left panel they are
plotted as a function of $eV/\Delta $ showing the different shapes and amplitudes of the individual
MAR processes for varying transmission. A small $T$ gives rise to narrow conductance spikes, whereas
a higher $T$ yields round maxima at voltages $V < 2\Delta/me$ and pronounced minima close to the
sub-multiple values $V = 2\Delta/me$. This behavior is clearly visible in the right panel where the
differential conductance is plotted as a function of the generalized order of the MAR process 
$m^\prime=2\Delta /eV$. For small $T$ the onset of the MAR processes is equidistant with spacing
$2\Delta/eV$ and their amplitudes
decrease very rapidly with $m^\prime$. For high transmission the position of the maxima is progressively shifted to higher
$m^\prime$ values, while the minima correspond approximately to integer values of $m^\prime$. 
The experimental data of Fig.~\ref{tdep} display a mixed character of high $T$ and low $T$ behavior,
because of the presence of the two extreme channels with $T_1=0.90$ and $T_2 = 0.108$. 
The value of the temperature dependent gap $\Delta(\Theta)$ can be
determined by the peak of the $m=1$ process, while the rest of the 
${\mathrm d}I/{\mathrm d}V$ is
dominated by the widely open channel $T_1$. 
In the left inset of Fig.~\ref{tdep} we plot the
position of the $m = 1$ maximum as a function of the temperature. 
Also shown are data taken on
different contact configurations of the same sample. 
The development of the peak 
position follows the BCS gap 
function $\Delta_{\mathrm BCS}(\Theta)$ which is plotted as a solid line in the same graph. 
We have verified for contacts with different conductances ranging from the tunnel regime to 
several $G_0$ that the $IV$s can be described by the same channel distribution (with restricted
accuracy due to less pronounced MAR features) up to the critical temperature. When exceeding
$\Theta_{\mathrm c}$ the $IV$ characteristics become linear with a slope corresponding to $D$
within 1\%.

\section{IVs of Al point contacts with external magnetic field}

\begin{figure}[btp]
\centerline{\psfig{file=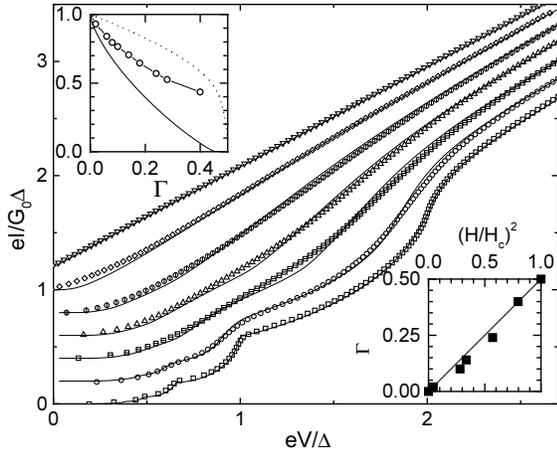,width=0.85\linewidth}}
\caption{ 
Measured (symbols) and calculated (lines) $IV$s of trace (d) of Fig.~\protect\ref{aluivs} with transmission
set $\{0.519, 0.253, 0.106\}$ for magnetic 
fields of $\mu_0H$ = 0, 2.1, 5.4, 5.9, 7.7, 9.1 and 10.9 mT (from bottom to top). The curves with
magnetic field are offset by 0.2 each.
 Left inset: pair amplitude $\Delta$ (dotted), spectral gap $\Omega$ (solid) (according to
 \protect\cite{skalski}) and position of the 
 $m = 1$ maximum of the corresponding calculated ${\mathrm d}I/{\mathrm d}V$s (symbols) in units of the
 gap parameter at zero field as a function of the pair breaking parameter $\Gamma$.
Right inset: Pair breaking parameter $\Gamma$ used for calculating the $IV$s and 
${\mathrm d}I/{\mathrm d}V$s
as a function of applied field (symbols) and prediction according to \protect\cite{maki,skalski} (line). 
The $IV$ at $H$ = 0 was measured at the beginning and the end of the series of measurements 
and the $\{ T_{\mathrm n}\}$ was found to be reproducible.
The critical field at the measuring temperature of $\Theta = 25~{\mathrm mK}$ was determined 
to be $\mu_0H_{\mathrm c} = 10.2$ mT.
}\label{bfield}
\end{figure}

Fig.~\ref{bfield} shows the evolution of the subgap structure with applied 
magnetic field. The traces are offset for clarity. In our experiment the field is applied 
perpendicular to the film plane. As the field size is
increased, the excess current is suppressed, the current steps are strongly rounded and the peak positions
are shifted to lower voltages. For fields larger than $\approx$ 5.0\,mT no clear
sub-multiple current steps are observable. When a field of 
$\mu_0H_{\mathrm c}\,=\,10.2\,{\mathrm mT}$  close to the 
bulk critical field of Al $\mu_0H_{\mathrm c,bulk}\,=\,9.9\,{\mathrm mT}$
is reached the $IV$ becomes again linear with a slope corresponding to the
sum $D$ of the transmissions determined in zero field. When lowering the field again to $H = 0$ we
recover the same subgap structure as before. When reversing the field direction the same $IV$ is
observed for the same absolute value of the field, which proves that there is no residual field
along the field axis. Effects of the earth magnetic field or spurious fields in different 
directions however cannot be excluded.

An external magnetic field suppresses superconductivity because it acts as an effective pair
breaking mechanism \cite{maki}. 
Since a magnetic field breaks the electron-hole symmetry, the $IV$s due to MAR are modified as
demonstrated by Zaitsev and Averin \cite{zaitsev}. Strictly speaking, the conduction channels 
could be different with and without magnetic field. 

A quantitative description of the influence of the magnetic field is difficult because of the
complicated shape of the samples. The point contact spectra are sensitive to the
superconducting properties at the constriction. Since the pair breaking parameter 
$\Gamma =\hbar/(2\tau_{\mathrm pb}\Delta)$ ($\tau_{\mathrm pb}$ is the pair-breaking time) due to an
external magnetic field is geometry dependent \cite{maki} a complete description 
needs to take into
account the exact shape of the sample on the length scale of the coherence length $\xi$. We
estimate for our Al films in the dirty limit $\xi =
(\hbar D_{\mathrm el}/2\Delta)^{1/2} = 280~{\mathrm nm}$ where 
$D_{\mathrm el} = v_{\mathrm F}l/3 = 0.042~{\mathrm m^2/s}$ 
is the electronic diffusion constant. 

Due to the
finite elastic mean free path $l\approx 65\,\mathrm{nm}$ (determined by the residual resistivity
ratio RRR $= R(300\,{\mathrm K})/R(4.2\,{\mathrm K}) \approx 4$) of the evaporated thin film the 
penetration depth is enhanced and it is comparable to the sample thickness. 
The fact that all 
signature of superconductivity is destroyed at the bulk critical
field indicates that the geometry of the sample does not play a dominant role, but that $\xi$ 
is the most important length scale.  We therefore describe the influence of the magnetic field along the lines of Skalski 
{\sl et al.} \cite{skalski} using a homogeneous $\Gamma$ given by the expression \cite{maki,belzig}
\begin{equation}
\Gamma = \frac {D_{\mathrm el}e^2\mu_0^2H^2w^2}{6\hbar\Delta}
\label{gamma}
\end{equation}
\noindent
where the effective width of the 
film $w = 280~{\mathrm nm} \simeq \xi$ is limited by the coherence length. 
Superconductivity is completely suppressed when $\Gamma = 0.5$.
In order to obtain the $IV$ curves for one
channel in an external magnetic field, the BCS density of states in the theory of Ref.
\cite{cuevas} is replaced by the corresponding expressions given in Refs. \cite{maki,skalski}
which include the effect of a pair breaking mechanism.

Contrary to the influence of higher temperatures, the magnetic field rounds the density 
of states and the Andreev reflection amplitude. 
The rounding is a consequence of the fact that the 
pair amplitude $\Delta$ and the
spectral gap  $\Omega$ of the density of states (i.e. the energy up to which the density of states is zero) 
differ from each other when time reversal symmetry is lifted.
The position of the $m = 1$ maximum of the 
${\mathrm d}I/{\mathrm d}V$ does not give an accurate estimation of $\Gamma$ 
and it is necessary to fit the whole $IV$. 
In the left inset we display the evolution (as a function of $\Gamma$) of $\Delta$, $\Omega$ and the position
of the maximum conductance of the $m = 1$ process of the contact whose $IV$s are shown in
Fig.~\ref{bfield}. The functional
dependence of the latter is not universal but depends on the distribution of transmissions. 

It turns out that the structure in the experimental data is more rounded
than in the calculated curves indicating the limitations of the model used here. 
A reasonable agreement between the experimental data and the 
model is found when determining $\Gamma$ such that the excess current is correctly described. 
The solid lines in Fig.~\ref{bfield} are calculated with the transmission ensemble determined
at zero field for values of $\Gamma$ given in the inset. These values correspond
nicely to the predicted quadratic behavior of Eq.~\ref{gamma}.
It was not possible to achieve better agreement between the measured and calculated $IV$s by 
altering the channel ensemble, supporting again that the
conduction channels are not affected by superconductivity, nor by its suppression \cite{suderow}.
 
We have demonstrated here, that it is possible to 
drive a particular contact reproducibly into the normal state and back to the superconducting state
without changing $\{ T_{\mathrm n}\}$. We stress the high stability of the setup
necessary for maintaining a particular contact stable during the measurement series.

\section{Conclusions}

We have reported measurements and the analysis of multiple Andreev reflection in
superconducting atomic contacts demonstrating that the
conduction channel ensemble of the smallest point contacts between Al electrodes consists of at 
least two, more often three channels.
We have verified that the channel ensemble remains unchanged when suppressing the 
superconductivity gradually by increasing the temperature or 
applying a 
magnetic field. This result strongly supports the expected equivalence of conduction channels in the 
normal and in the superconducting state and agrees with the quantum chemical picture of
conduction channels. The latter suggests that the conduction channels are determined by the band
structure of the metal and therefore their transmissions vary significantly only on the scale of
several $e$V. Superconductivity, which opens a spectral gap for
quasiparticles of the order of only $\simeq$ m$e$V does not modify the channels and is therefore a
useful tool to study them.

We thank C. Strunk and W. Belzig for valuable discussions. This work was supported in part by the 
Deutsche Forschungsgemeinschaft (DFG), Bureau
National de M\'etrologie (BNM), and the Spanish CICYT.


\begin{thebibliography}{99}


\bibitem{jancuracao} J.M. van Ruitenbeek, in "Mesoscopic Electron Transport",
 L.L. Sohn, L.P. Kouwenhoven, and G. Sch\"on eds., Amsterdam, 1997, and references 
therein.
\bibitem{landauer} R. Landauer, IBM J. Res. Dev. 1 (1957) 223; Philos. M. 21 (1970) 863.
\bibitem{olesen}  L. Olesen {\sl et al.}, Phys. Rev.
Lett. 72 (1994) 2251.
\bibitem{krans95}  J.M. Krans {\sl et al.}, Nature  375 (1995) 767.
\bibitem{costa-krämer}  J.L. Costa-Kr\"{a}mer, N.P. Garc\'{\i}a-Mochales,
and P.A. Serena, Surface Science 342 (1995) L1144.
\bibitem{yansoncomment}A.I. Yanson and J.M. van Ruitenbeek, Phys. Rev. Lett. 79 (1997) 2157. 
\bibitem{aluprl} E. Scheer {\sl et al.}, Phys. Rev. Lett. 78 (1997) 3535.
\bibitem{arnold87}  L.B. Arnold, Journal of Low Temp. Phys. 68 (1987) 1.
\bibitem{averin95}D. Averin and A. Bardas, Phys. Rev. Lett. 75 (1995) 1831.
\bibitem{cuevas} J.C. Cuevas, A. Levy Yeyati, and A. Mart\'{\i}n-Rodero, 
Phys. Rev. B 54 (1996) 7366.
\bibitem{shumeiko} E.N. Bratus {\sl et al.}, Phys. Rev. B 55
(1997) 12666.
\bibitem{note1}  The four works \cite{arnold87,averin95,cuevas,shumeiko} deal
through different approaches with the same physics and provide essentially
the same results for the IV. We have used the numerical results provided by
\cite{cuevas} in order to draw the inset of Fig. \ref{aluivs} and to perform the fits.
\bibitem{carlosprl}J.C. Cuevas, A. Levy Yeyati, A. Mart\'{\i}n-Rodero, Phys.
Rev. Lett. 80 (1998) 1066.
\bibitem{scheernature} E. Scheer {\sl et al.}, Nature 394 (1998)
154.
\bibitem{aupaper}E. Scheer, W. Belzig, Y. Naveh, D. Esteve, C. Urbina, and M.H.
Devoret, in preparation.
\bibitem{helko} H.E. van den Brom, J.M. van Ruitenbeek, Phys. Rev. Lett. 82 (1999) 1526
\bibitem{ludophcf} B. Ludoph {\sl et al.}, Phys.
Rev. Lett. 82 (1999) 1530.
\bibitem{ludophtp} B. Ludoph, J.M. van Ruitenbeek, Phys. Rev. B 59 (1999) 12290. 
\bibitem{BCS}J. Bardeen, L.N. Cooper, and J.R. Schrieffer, Phys. Rev. 108 (1957) 1175.
\bibitem{beenakkerprb} C.W.J. Beenakker, Phys. Rev. B 46 (1992) 12841.
\bibitem{averinprbrc} A. Bardas and D.V. Averin, Phys. Rev. B 56 (1997) R8518.
\bibitem{carlosdiss}  A. Martin-Rodero, A. Levy Yeyati, J.C. Cuevas, Superlattices and Microstructures,
       25 (1999) 925.
\bibitem{carlosunpub} J.C. Cuevas, unpublished.
\bibitem{klapwijk}  T.M. Klapwijk, G.E. Blonder and M. Tinkham,
Physica 109\&110B (1982) 1657.
\bibitem{hurd}  M. Hurd, S. Datta, and P.F. Bagwell, Phys.
Rev. B 54 (1996) 6557.
\bibitem{burstein}  B.N. Taylor and E. Burstein, Phys. Rev. Lett. 10 (1963) 14.
\bibitem{flensberg89}  K. Flensberg, and J. Bindslev Hansen, Phys. Rev B 40 (1989) 8693.
\bibitem{kleinsasser}  A.W. Kleinsasser {\sl et al.}, Phys. Rev. Lett. 72 (1994) 1738.
\bibitem{jansaclay} J.M. van Ruitenbeek {\sl et al.}, 
Rev. Sci. Inst. 67 (1996) 108.
\bibitem{glattli}  D.C. Glattli {\sl et al.}, J. Appl. Phys. 81 (1997) 7350.
\bibitem{vion}  D. Vion {\sl et al.}, J. Appl. Phys. 77 (1995) 2519.
\bibitem{maki} K. Maki in "Superconductivity", vol. 2, R.D. Parks ed., M. Dekker (New York,1969).
\bibitem{zaitsev}A.V. Zaitsev and D.V. Averin, Phys. Rev. Lett. 80 (1998) 3602.
\bibitem{skalski} S. Skalski, O. Betbeder-Matibet, and P.R. Weiss, Phys. Rev. 136 (1964) A1500.
\bibitem{belzig} W. Belzig, C. Bruder, and G. Sch\"on, Phys. Rev. B 54 (1996) 9443.
\bibitem{suderow} In a recent work by Suderow {\sl et al.} 
on long neck contacts of Pb prepared in an STM a strongly enhanced
critical field was observed. This was explained by the varying sample width on the length scale 
of $\xi$, which enables superconductivity in the long neck although the bulk critical field is
exceeded. In Al $\xi$ is much larger than in Pb and therefore this effect appears to be 
negligible for the present work. However, the observed rounding of the MAR structures in our
samples might  be due to a similar mechanism. H. Suderow {\sl et al.}, cond-mat/9907236.
\end{thebibliography}
\end{document}